\pdfoutput=1

\documentclass[11pt]{article}
\usepackage{xcolor} 
\usepackage[]{EMNLP2023}
\usepackage{EMNLP2023}
\usepackage{graphicx}
\usepackage{times}
\usepackage{enumitem}
\usepackage{multirow}
\usepackage{url}
\usepackage{latexsym}
\usepackage{booktabs}
\usepackage{amsmath, amssymb}

\newcommand{\distance}{6pt}
\usepackage[T1]{fontenc}

\usepackage[utf8]{inputenc}
\usepackage{microtype}

\usepackage{inconsolata}
\usepackage{xspace} 
\newcommand{\bench}{\textsc{PsyScam}\xspace} 
\title{PsyScam: A Benchmark for Psychological Techniques in Real-World Scams}

\author{
\small
    Shang Ma\textsuperscript{1}, 
    Tianyi Ma\textsuperscript{1}, 
    Jiahao Liu\textsuperscript{2}, 
    Wei Song\textsuperscript{1},\\ 
    \small
    \textbf{
    Zhenkai Liang\textsuperscript{2},
    Xusheng Xiao\textsuperscript{3}\textsuperscript{$\dagger$},
    Yanfang Ye\textsuperscript{1}\textsuperscript{$\dagger$}}\\
\small
    \textsuperscript{1}University of Notre Dame, 
    \textsuperscript{2}National University of Singapore,
    \textsuperscript{3}Arizona State University \\
\small
    \textsuperscript{$\dagger$}Corresponding Authors
    \\
\small
    \texttt{\{sma5,tma2,wsong8,yye7\}@nd.edu},
    \\ 
\small
    \texttt{\{jiahao99,liangzk\}@comp.nus.edu.sg},
        \texttt{xusheng.xiao@asu.edu}
}

\begin{document}
\maketitle
\begin{abstract}

\blue{
Over the years, online scams have grown dramatically, with nearly $50\%$ of global consumers encountering scam attempts each week.
These scams cause not only significant financial losses to individuals and businesses, but also lasting psychological trauma, largely due to scammers' strategic employment of psychological techniques (PTs) to manipulate victims.
Meanwhile, scammers continually evolve their tactics by leveraging advances in Large Language Models (LLMs) to generate diverse scam variants that easily bypass existing defenses.}

\blue{
To address this pressing problem, we introduce \bench, a benchmark designed to systematically capture the PTs employed in real-world scam reports, and investigate how LLMs can be utilized to generate variants of scams based on the PTs and the contexts provided by these scams.
Specifically, we collect a wide range of scam reports and ground its annotations of employed PTs in well-established cognitive and psychological theories.
We further demonstrate LLMs' capabilities in generating through two downstream tasks: scam completion, and scam augmentation.
Experimental results show that \bench~presents significant challenges to existing models in both detecting and generating scam content based on the PTs used by real-world scammers.
Our code and dataset are available\footnote{\url{https://github.com/KiteFlyKid/PsyScam}}.}

\end{abstract}

 \section{Introduction}
\begin{figure}[t]
    \centering
    \includegraphics[width=\columnwidth]{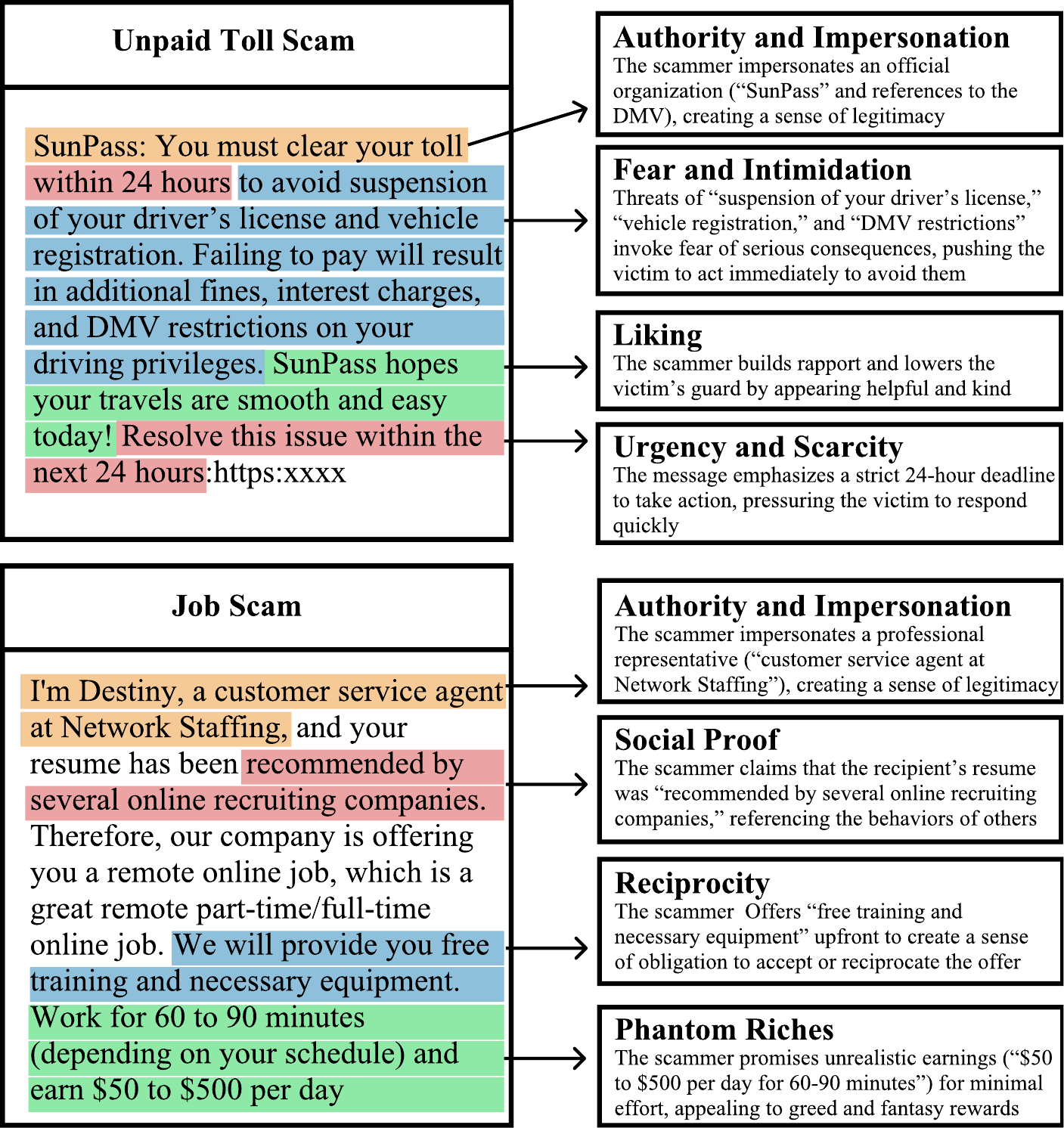}
    \caption{Psychological techniques in prevalent scams.}
    \label{fig:goldenexample}
\end{figure}
Online scams have become a global epidemic, causing severe financial and psychological harm to individuals and organizations. According to the U.S. Federal Trade Commission (FTC), consumers reported over $5.7$ billion in losses to fraud in 2024 alone, marking a 125\% increase from the previous year~\cite{ftcScamStats}. Similarly, Singapore’s government recorded $929.6$ million in scam-related losses in 2024, reflecting a 41\% year-over-year rise~\cite{sgScamStats}. In Europe, Nasdaq Verafin estimates that $103.6$ billion in illicit funds linked to fraud and scams flowed through the financial system in 2024~\cite{euScamStats}. These alarming figures highlight the escalating scale and impact of scams worldwide.


\blue{
Unlike traditional cyber attacks that exploit technical methods, scammers employ psychological techniques (PTs) to manipulate victims~\cite{montanez2022cyber,longtchi2024internet}, resulting not only in significant financial losses but also lasting psychological trauma. As illustrated in \autoref{fig:goldenexample}, scammers can effectively combine multiple PTs in their messages to carry out various scams, such as unpaid toll~\cite{TollScamFBI,TollScamFTC} and job scams~\cite{jobScam}. Furthermore, scammers continuously evolve their tactics to evade detection and bypass existing deep learning-based detectors, which are highly sensitive to out-of-distribution (OOD) samples. For example,  a minor variation in the wording of scam messages can target different demographics~\cite{federalReserve}. To make matters worse, with the advancements of Large Language Models (LLMs), scammers now have new opportunities to produce a wider array of scam variants that can easily circumvent existing detection mechanisms,  significantly deepening the societal harms caused by these scams.     }


\textbf{Why this benchmark matters:}
Recent efforts to benchmark scams face two key limitations. First, existing work relies on synthetic data generated by large language models (LLMs)\cite{yang2025fraud,roy2024chatbots}, which often fail to reflect the 
 context in actual scammer-victim interactions. 
 Second, although a few studies leverage real-world scam data, they generally restrict their scope to specific scam types like smishing~\cite{timko2024smishing} or phishing emails~\cite{chakraborty2024detoxbench}, 
 and do not explicitly model the psychological aspects of the scams.
 As a result, there is a lack of comprehensive benchmarks that combine real scam data with annotations of underlying PTs.

To address these limitations, we introduce \bench, \textit{the first benchmark designed to systematically capture and evaluate the PTs embedded in real-world scam incidents}. 
Specifically, we first collect scam reports from six scam reporting platforms that cover a diverse range of scam incidents, including online or offline, delivered via email or SMS, crypto-related or not, and originating from both the U.S. and other global regions.
Second, building on foundational cognitive and psychological theories, we construct a taxonomy of nine PTs that frequently appear in scams. We then employ a human–LLM collaborative annotation pipeline to efficiently and effectively label PTs present in each report.
Furthermore, to demonstrate the utility of \bench, we define three representative downstream tasks: (1) \textit{PT Classification}, which maps scam texts to their corresponding PTs; (2) \textit{Scam Completion}, which predicts scam texts aligned with given PTs; and (3) \textit{Scam Augmentation}, which rewrites scam texts to incorporate new PTs.  
We conduct extensive experiments using a variety of baselines, including traditional models and LLMs, to illustrate how \bench~presents meaningful challenges and opportunities for advancing cybersecurity research.  
Our key contributions can be summarized as below:
\begin{itemize}[noitemsep, topsep=1pt, partopsep=1pt, listparindent=\parindent, leftmargin=*]
    \item \textbf{Novel Benchmark for Psychological Techniques in Scams.}
We present \bench, the first benchmark to capture PTs in real-world scam reports, addressing a critical gap in scam analysis by grounding PT annotations in authentic scammer-victim interactions. 
    \item \textbf{Human–LLM Collaborative Annotation at Scale.}
We develop a scalable annotation framework that combines the interpretive strength of humans with the extraction capabilities of large language models, enabling high-quality PT labeling across thousands of real scam reports. 
    \item \textbf{Comprehensive Tasks and Evaluation.} We define three representative downstream tasks (i.e., PT classification, scam completion, and scam augmentation), and conduct extensive experiments using both traditional classifiers and state-of-the-art LLMs, demonstrating how \bench~helps advance scam detection and generation research. 
\end{itemize}

\section{Related Work}

\subsection{Empirical Studies of Scams}
Previous research has empirically studied diverse scam types individually, utilizing various methodologies to gain insights into scammers' strategies and operations. For instance, scambaiter~\cite{park2014scambaiter} employed honeypot advertisements on Craigslist to attract scammers involved in advanced fee scams (i.e., Nigerian scam), interacting directly to analyze scammers' operational patterns. 
Similarly, researchers developed Twitter-based honeypots, automatically engaging cryptocurrency-based technical support scammers to systematically study their tactics~\cite{acharya2024conning}. 
A different research effort focused on cryptocurrency investment scams, employing large-scale web crawling to collect and analyze deceptive websites~\cite{muzammil2025poorest}.  
Differing from these narrowly scoped studies, our benchmark covers a broad spectrum of scam types, emphasizing communication strategies used by scammers to seduce and engage victims.

\subsection{Psychological Factors in Scams}
Parallel to these technical efforts in characterizing scams, a growing body of academic work examines the psychological and social engineering techniques that make scams effective.
Nelms et al.~\cite{nelms2016towards} studied software download attacks, highlighting techniques that scammers use to capture user attention, deceive, and persuade victims. Van der Heijden and Allodi~\cite{van2019cognitive} applied principles from Cialdini's persuasion theory~\cite{cialdini2007influence} to analyze phishing emails, providing a cognitive framework for prioritizing and mitigating phishing threats based on psychological manipulation tactics. Extending cyber threat frameworks such as MITRE ATT\&CK~\cite{attck}, Montañez and Xu~\cite{montanez2022cyber} proposed a cyber social engineering kill chain, further detailed by Longtchi et al.~\cite{longtchi2024internet}, delineating stages of psychological manipulation employed by scammers.
Building on these foundational insights, our work defines explicit PTs employed by scammers and, importantly, contributes a publicly available benchmark dataset annotated with PTs to support future research in scam analysis and mitigation.


\section{The PsyScam Benchmark}

\subsection{Dataset Collection}

\begin{table}[t]

\resizebox{\columnwidth}{!}{%
\begin{tabular}{@{}llll@{}}
\toprule
Name                    & Region  & Size      & Focus                                         \\ \midrule
BBB Scam Tracker        & US      & 10,000   & All types of scam \\
Scamsearch              & World   & 204,560 & All types of scam              \\
Crypto Scam Tracker     & US (CA) & 291       & Cryptocurrency-related scams                  \\
Investment Scam Tracker & US (WI) & 34        & Investment and financial scams                \\
Scam-Tracking Map       & US      & 1,000      & Geolocated scams                     \\
SmishTank               & World   & 20,295    & Smishing (SMS phishing) scams                 \\ \bottomrule
\end{tabular}
}

\caption{Scam datasets}
\label{tab:dataset_sources}
\end{table}

 \autoref{tab:dataset_sources} summarizes our dataset collected from six prominent scam report platforms. Scam report platforms enable users to directly upload detailed scam experiences, thus providing firsthand reports across various communication channels, including email, SMS, phone calls, and social media. For instance, \autoref{fig:scam_report} shows a typical scam report submitted to the Better Business Bureau (BBB), illustrating the depth and richness of information available from these user-generated reports. Our dataset covers a comprehensive range of scam types, such as employment scams, job scams, investment scams, cryptocurrency scams, phishing, etc. 

As no open-source datasets currently exist for these scam reporting platforms, and direct open access is typically restricted, we develop custom web crawlers tailored to each platform to systematically collect these reports.
Scam reports are sometimes repeatedly submitted by the same victim or exhibit sudden spikes of similar incidents within a specific time frame. 
To address these duplicated reports, we generate embeddings of each crawled scam report using BERT and remove those with cosine similarity above a certain threshold. Given the abundant amount of data, we employ an aggressive similarity threshold of $0.8$, effectively removing most duplicates. Additionally, we exclude excessively brief scam reports, such as random complaints unlikely to contain relevant PTs, by filtering out entries below the 20th percentile in length ($31$ words). 


\begin{table*}[t]
\resizebox{\linewidth}{!}{%
\begin{tabular}{lll}
\hline
PTs                   & Description                                                & Example                                                                                                                                         \\ \hline
Authority and Impersonation             & Tend to obey authorities and credible individuals                                    & \begin{tabular}[c]{@{}l@{}}``Person claimed to be calling for Finance America, \\ claiming our home warranty was expired"\end{tabular}           \\
Phantom Riches        & Visceral triggers of desire that override rationality      & \begin{tabular}[c]{@{}l@{}}``Your phone Number was randomly selected from \\ the US database and you have won 18,087.71"\end{tabular}            \\
Fear and Intimidation & Fear of loss and penalties                                 & ``You will be arrested!''                                                                                                                         \\
Liking                & Preference for saying ``yes'' to people they like            & \begin{tabular}[c]{@{}l@{}}``I am always available to help, and it's my pleasure \\ to answer any questions you may have"\end{tabular}           \\
Urgency and Scarcity  & Sense of urgency and scarcity assign more value to items   & ``We are currently in urgent need of 100 employees''                                                                                              \\
Pretext and trust     & Tendency to trust credible individuals                     &
\begin{tabular}[c]{@{}l@{}} ``This is an urgent message for [MY NAME]. I'm calling regarding \\a complaint scheduled to be filed out of [Our County Name]''    \end{tabular}                                                                                                  \\
Reciprocity           & Tendency to feel obliged to repay favors from others       & \begin{tabular}[c]{@{}l@{}}``We will send you a check to purchase equipment \\ such as new apple laptop and iphone 14 and software''\end{tabular} \\
Consistency           & Tendency to behave consistently with past behaviors        & \begin{tabular}[c]{@{}l@{}}Starts with small asks (fill a form) and \\ escalate to big asks (invest money)\end{tabular}                         \\
Social Proof          & Tendency to refer majority's behavior to guide own actions & \begin{tabular}[c]{@{}l@{}}``Your resume has been recommended by \\ many online recruitment companies''\end{tabular}                              \\ \hline
\end{tabular}
}
\caption{Psychological techniques.}
\label{tab:PTs}
\end{table*}


\subsection{Psychological Techniques} 
To model psychological techniques present in scams, we compile a taxonomy (shown in \autoref{tab:PTs}) based on elements from well-established psychological and behavioral theories, chosen to reflect the persuasive strategies frequently exploited in real-world scams. 
These techniques are grounded in decades of empirical research on human influence and decision-making.

Specifically, six techniques—\textit{Authority and Impersonation}, \textit{Reciprocity}, \textit{Consistency}, \textit{Social Proof}, \textit{Liking}, and \textit{Urgency and Scarcity}—are derived from Cialdini’s Principles of Persuasion~\cite{cialdini2007influence}. These principles explain how individuals are influenced by perceived authority, obligations to return favors, social norms, and time-sensitive pressure. 

Two additional techniques—\textit{Fear and Intimidation} and \textit{Phantom Riches}, which are drawn from Prospect Theory~\cite{kahneman2013prospect}, explain how individuals make decisions based on perceived gains and losses. Scammers often leverage fear of loss (e.g., threats of legal action or account suspension) or exaggerated gain (e.g., guaranteed investment returns) to manipulate decision-making under emotional pressure.

The final technique—\textit{Pretext and Trust}—is based on the Elaboration Likelihood Model (ELM)~\cite{petty2012communication}, particularly the peripheral route of persuasion. This occurs when individuals rely on superficial cues, such as familiarity, friendliness, or informal tone, rather than critical thinking. Scammers frequently exploit this through stolen personal information or misleading personal references to build false rapport.


\subsection{Human-LLM Collaborative Annotation}
\label{subsec:dataannotation}

\noindent\textbf{Motivation.} 
Scam reports are often written by everyday users and may include grammar errors, emotionally charged language, irrelevant details, or ambiguous phrasing, which makes it difficult to accurately extract the underlying PTs. This variability presents a major challenge for annotation: while expert human labeling can be accurate, it is also error-prone and time-consuming at scale.
An alternative is to use LLMs to automate annotation. However, LLMs are prone to hallucination, generating incorrect or overly confident predictions, and often associate scam reports with irrelevant or excessive PT labels. To balance precision and efficiency, we adopt a two-stage collaborative framework that leverages the strengths of both humans and LLMs.

\noindent\textbf{LLM as Extractor.} 
Specifically,  we first use few-shot prompting to instruct the LLM to extract candidate PTs and the corresponding supporting texts from scam reports. As shown in \autoref{tab:promptAnnotation} in \autoref{appendix:Prompt}, the prompt clearly defines the task, specifies format requirements, and provides concrete examples to guide the LLM's behavior. The LLM is explicitly instructed to avoid guessing and to return a structured JSON dictionary mapping each predicted PT to a supporting excerpt from the report. 

\blue{
To build the few-shot prompt, we curate examples by manually labeling 20 reports for each PT, selected from the high-quality and diverse BBB dataset. 
Our error analysis shows that the LLM-based extractor successfully captures all key scam-related texts, achieving a recall of $0.77$ and a precision of $0.59$ in correctly classifying the PTs of the extracted text.   
These results highlight both the effectiveness of the extractor in identifying relevant content and the necessity of the subsequent human verification stage in our framework.}

\noindent\textbf{Human as Verifier.} 
In the second stage, human annotators review the LLM’s output to ensure correctness. In practice, we observe that the LLM tends to extract more PTs than necessary. However, it rarely misses truly relevant PTs.   Therefore, human annotators are able to primarily focus on verifying whether the extracted text accurately reflects the assigned PTs, rather than identifying PTs from scratch. This significantly reduces annotation time and effort. 


\blue{
The human annotation is performed by three graduate researchers in computer science.   To ensure consistent understanding of PTs and minimize ambiguity, we 1) conduct a practice annotation prior to the formal annotation and 2) create a detailed reference table defining each PT clearly and providing sufficient examples to distinguish fuzzy boundaries. Each disagreement is resolved through discussion or adjudication by a third reviewer. }

\begin{figure}[t]
    \centering
    \includegraphics[width=\columnwidth]{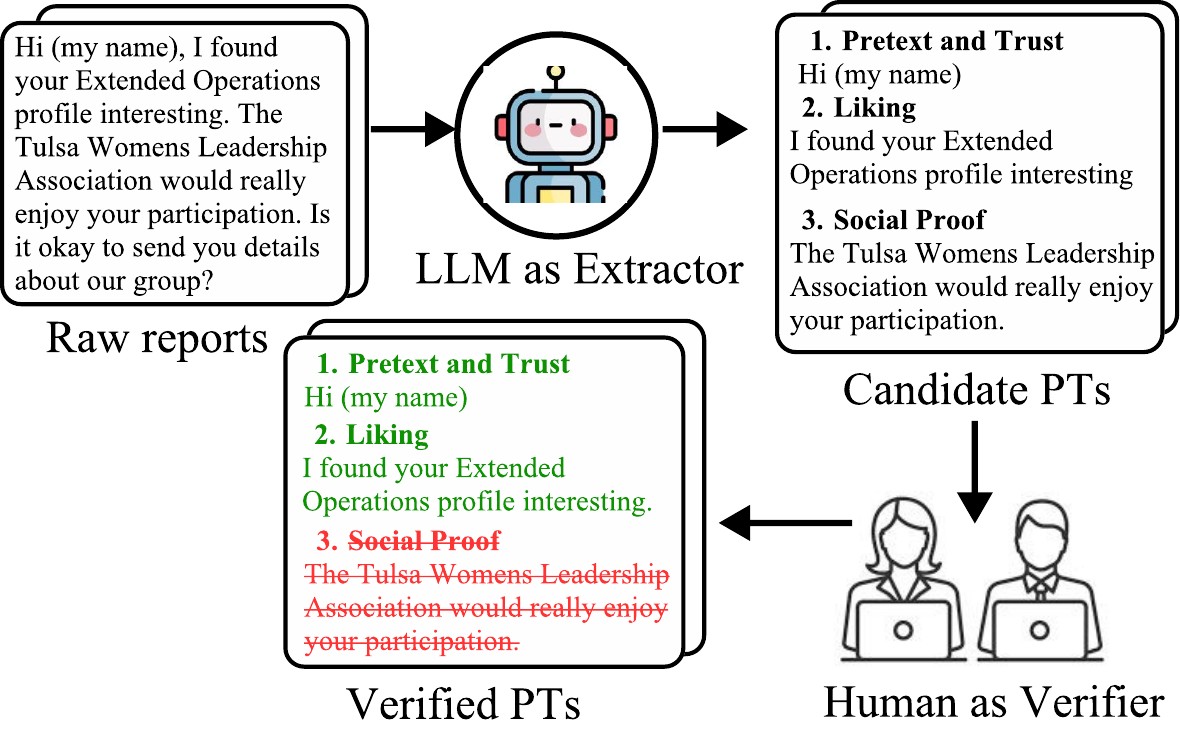}
    \caption{Human-LLM collaborative annotation.}
    \label{fig:label_process}
\end{figure}

\subsection{Dataset}

The full dataset encompasses all collected scam reports detailed in \autoref{tab:dataset_sources}. We provide open access to this comprehensive dataset for research purposes. Additionally, these scam reports are also structured and standardized by following the STIX 2.1~\cite{stix} specification, facilitating interoperability and ease of use in cybersecurity analyses.  
o demonstrate the utility of \bench, we further curated three datasets:

\noindent\textbf{D0:} D0 is derived from all sources listed in \autoref{tab:dataset_sources}. We clean the data by removing informationless reports based on text length (e.g., about $50\%$ of reports from the ScamSearch dataset contain six or fewer words). In total, D0 contains $16,658$ reports.  

\noindent\textbf{D1:} D1 contains $1126$ scam reports manually annotated with PTs. To optimize annotation quality and efficiency, we specifically choose and annotate reports from  BBB Scam Tracker,  Crypto Scam Tracker, and Investment Scam Tracker, as these sources typically offer detailed and high-quality descriptions essential for accurate annotation.  

\noindent\textbf{D2:} D2 is a subset of D1, consisting of $730$ scam reports that specifically include messages directly from scammers.  Scam reports often blend with messages directly quoted from scammers (e.g., \autoref{fig:scam_report}), victim narratives (e.g., \autoref{fig:scam_report_narrative}), or combinations thereof (e.g., \autoref{fig:scam_report_mix}).
We employed a two-step approach to construct this dataset. First, we manually annotated $200$ reports to train a binary classification model based on RoBERTa, achieving a robust F1 score of $97.45\%$ in classifying whether a scam report contains scammer quotations. Subsequently, we applied this model on the D1 dataset, keeping reports containing scammer messages, resulting in the final D2 dataset. 


\begin{figure}[t]
    \centering
    \includegraphics[width=\columnwidth]{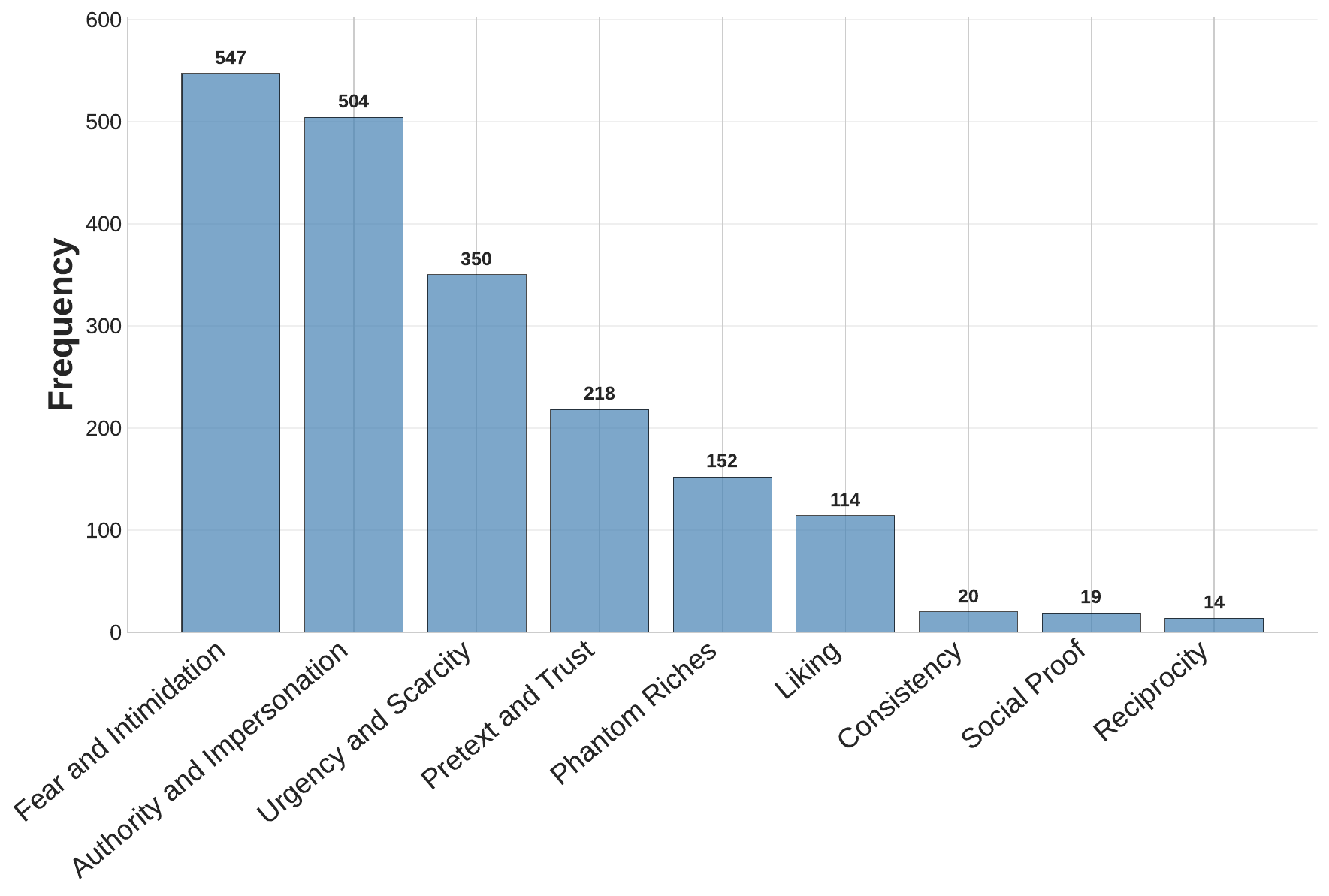}
    \caption{The distribution of PTs in D2. }
    \label{fig:ptdistribution}
\end{figure}

\blue{
\autoref{fig:ptdistribution} illustrates the distribution of PTs in D2.  
Certain PTs, such as \textit{Authority and Impersonation} (547 cases), occur far more frequently than others like \textit{Reciprocity} (14 cases) and \textit{Social Proof} (19 cases).  
This imbalance reflects genuine scammer strategies rather than data collection bias: scammers tend to favor simpler, more effective techniques (e.g., Authority and Impersonation) over more complex ones (e.g., Reciprocity, Consistency). }


\begin{table}[t]

\resizebox{\columnwidth}{!}{%
\begin{tabular}{@{}llll@{}}
\toprule
        & \multicolumn{1}{c}{\textbf{PT Classification}}                             & \multicolumn{1}{c}{\textbf{Scam Completion}}                                    & \multicolumn{1}{c}{\textbf{Scam Augmentation}}                                  \\ \midrule
Input   & Scam Text                                                                  & Scam Text, PT                                                                   & Scam Text, PT                                                                   \\
Output  & PT                                                                         & Scam Text                                                                       & Scam Text                                                                       \\
Dataset & D1                                                                         & D2                                                                              & D2                                                                              \\
Metrics & \begin{tabular}[c]{@{}l@{}}Accuracy, Recall, \\ Precision, F1\end{tabular} & \begin{tabular}[c]{@{}l@{}}ROUGE, BertScore, \\ BLEU, SR\end{tabular} & \begin{tabular}[c]{@{}l@{}}ROUGE, BertScore, \\ BLEU, SR\end{tabular} \\ \bottomrule
\end{tabular}
}
\caption{Tasks overview.}
\label{tab:tasks}
\end{table}

\section{Task Design}
To demonstrate the utility of our benchmark, we design three representative downstream tasks, which are defined as follows:

\subsection{PT Classification}

\noindent\textbf{Task Setting}:
This task aims to automatically identify which PTs are used in a scam report. Since scammers often exploit multiple PTs simultaneously, this is framed as a multi-label classification task, classifying an instance (a scam report) to multiple labels (i.e., PTs) simultaneously.  
Automatically identifying the PTs in scam reports is essential for understanding scam strategies at scale and helping platforms monitor and respond to emerging threats. This task is also the base of further analysis.  We use dataset D1 to evaluate this task.

\noindent\textbf{Evaluation Metrics}:
We employ standard metrics for classification tasks, including accuracy, recall, precision, and F1 score for comprehensive performance assessment.

\subsection{Scam Completion}
\label{subsec:scamcompletion}
\noindent\textbf{Task Setting}:
This task simulates how scammers might continue their communications. It can help train or evaluate systems that aim to detect scams before victims are fully manipulated~\cite{scamshield}. 
Specifically, we evaluate whether an LLM can continue a scam message in a way that exhibits a given set of PTs.  
We provide the LLM with the beginning of a real scam message ask it to generate a plausible continuation.  We use dataset D2 to evaluate this task.  


\noindent\textbf{Evaluation Metrics}:
Our goal is to ensure the scam message preserves the original facts, meaning, and expresses the same PTs as the original scam message. To this end, we employ four metrics: \textit{ROUGE}, \textit{BLEU}, \textit{BERTScore} and \textit{success rate} (SR).

\textit{ROUGE} and \textit{BLEU} capture syntactic similarity by measuring n-gram overlap between the generated and original texts, helping assess whether key factual details (e.g., names, numbers, deadlines) are retained.

\textit{BERTScore} computes token-level similarity in embedding space using a pretrained language model. This metric helps assess whether the generated content conveys the same meaning as the original message, even if the phrasing differs. This helps because LLM often generates text with diversified words but preserves the same semantic meanings, rendering n-gram metrics less ineffective. 


\textit{SR} measures whether the generated message reflects the same PTs as the original scam message. It is defined as: 
\begin{equation}
\text{SR} = \frac{1}{N} \sum_{i=1}^{N} \mathbb{1}\left[ \text{PT}_{\text{pred}}^{(i)} = \text{PT}_{\text{true}}^{(i)} \right] 
\end{equation}, 
where \( N \) is the number of samples, and \( \mathbb{1}[\cdot] \) is the indicator function that evaluates to $1$ if the predicted set of PTs exactly matches the ground-truth set for sample \( i \), and $0$ otherwise.  We calculate SR by applying the LLM-based annotator described in Section~\ref{subsec:dataannotation} to extract PTs from the generated scam message and compare them with the original PTs.

\subsection{Scam Augmentation}
\noindent\textbf{Task Setting:}  
\blue{This task focuses on augmenting the dataset by generating scam messages conditioned on specified PTs.  
Given the imbalanced distribution of PTs in real-world scams, we leverage LLMs to generate synthetic examples, thereby enhancing the dataset and enabling the training of more robust classification models~\cite{yang2025fraud}.  
Our preliminary study shows that such augmentation improves model performance, particularly for less frequent PTs.  
Specifically, in this task we instruct an LLM to rewrite an existing scam message to incorporate a new PT that was not originally present.   
The prompt used for this task is provided in \autoref{tab:promptAugmentation} in \autoref{appendix:Prompt}.  
We use dataset D2 as the source for augmentation.  }


\noindent\textbf{Evaluation metrics}: 
We have a similar goal for this task as for Scam Completion: the generated output must preserve the original facts and meaning while incorporating the selected PTs.
Therefore, we apply the same metrics as in Scam Completion: ROUGE, BLEU, BERTScore, and SR. The SR here checks whether the added PT was successfully reflected in the rewritten message.


\section{Task Evaluation}

\begin{table}[t]

\resizebox{\columnwidth}{!}{%
\begin{tabular}{@{}lllll@{}}
\toprule
              & \multicolumn{1}{c}{\textbf{Accuracy}} & \multicolumn{1}{c}{\textbf{Recall}} & \multicolumn{1}{c}{\textbf{Precision}} & \multicolumn{1}{c}{\textbf{F1}} \\ \midrule
RoBerta-Based  & \textbf{0.4351}                       & \textbf{0.8987}                     & \textbf{0.8374}                        & \textbf{0.8669}                 \\
Bert-Based     & 0.4156                                & 0.9024                              & 0.8293                                 & 0.8643                          \\
SVM           & 0.3889                                & 0.8815                              & 0.8339                                 & 0.8569                          \\
Random Forest & 0.3953                                & 0.8872                              & 0.8288                                 & 0.8569                          \\
GPT-4.1-mini   & 0.2247                                & 0.7203                              & 0.8300                                 & 0.7713                          \\
Qwen3-30B          & 0.0137                                & 0.4872                              & 0.5708                                 & 0.5257                          \\ \bottomrule
\end{tabular}
}
\caption{Experimental results of PT classification on 6 baseline models.}
\label{tab:pt_classification}
\end{table}

\begin{table*}[t]
\small

\centering

\begin{tabular}{@{}llllllll@{}}
\toprule
                            & \textbf{Model} & \textbf{ROUGE-1} & \textbf{ROUGE-2} & \textbf{ROUGE-L} & \textbf{BLEU}   & \textbf{BERT}   & \textbf{SR}     \\ \midrule
\multirow{8}{*}{Completion} & GPT-4.1        & 0.1873           & 0.0258  & \textbf{0.1679}  & 0.0204          & 0.8137          & 0.3121          \\
                            & GPT-4.1-mini   & 0.1690           & 0.0178           & 0.1522           & 0.0178          & 0.8138          & \textbf{0.3376} \\
                            & GPT-4o         & 0.1821           & 0.0246           & 0.1642           & \textbf{0.0219} & \textbf{0.8181} & 0.2992          \\
                            & Gemini-2.0     & 0.1864           & \textbf{0.0269}           & 0.1669           & 0.0212          & 0.8088          & 0.2591          \\
                            & Grok-3         & 0.1727           & 0.0183           & 0.1514           & 0.0183          & 0.8086          & 0.2501          \\
                            & Grok-3-mini    & 0.1672           & 0.0165           & 0.1444           & 0.0161          & 0.8152          & 0.3322          \\
                            & Qwen-3-30B     & 0.1706           & 0.0195           & 0.1537           & 0.0188          & 0.8152          & 0.3156          \\
                            & Llama-3-70B    & \textbf{0.1947}  & 0.0249           & 0.1654           & 0.0199          & 0.8149          & 0.2811          \\ \midrule
\multirow{8}{*}{Augmentation}               & GPT-4.1        & 0.6400           & 0.4902           & 0.6208           & 0.4411          & 0.9180          & \textbf{0.8831} \\
                            & GPT-4.1-mini   & 0.7056           & 0.5659           & 0.6950           & 0.5211          & 0.9312          & 0.7842          \\
                            & GPT-4o         & 0.6064  & 0.4536  & 0.5940  & 0.4096 & 0.9096 & 0.7841          \\
                            & Gemini-2.0     & 0.6283           & 0.4809           & 0.6141           & 0.4477          & 0.9191          & 0.7277          \\
                            & Grok-3         & 0.7462           & 0.6512           & 0.7416           & 0.5827          & 0.9406          & 0.8761          \\
                            & Grok-3-mini    & 0.7353           & 0.6388           & 0.7279           & 0.5912          & 0.9323          & 0.8113          \\
                            & Qwen-3-30B     & \textbf{0.7867}  & \textbf{0.7253}  & \textbf{0.7824}  & \textbf{0.6628} & \textbf{0.9451} & 0.6547          \\
                            & Llama-3-70B    & 0.6914           & 0.6053           & 0.6876           & 0.5181          & 0.9294          & 0.8336          \\ \bottomrule
\end{tabular}
\caption{Experimental results of scam completion and scam augmentation.}
\label{tab:scam_generation}
\end{table*}

\subsection{PT Classification}
\noindent\textbf{Evaluation Setting.} 
We employ multiple baseline models categorized into three distinct types for comparison: traditional machine learning approaches (TF-IDF encoding~\cite{tfidf} + SVM~\cite{svm} and Random Forest~\cite{rf}) evaluated through 10-fold cross-validation; BERT-based models (BERT-based uncased~\cite{devlin2019bert}, RoBERTa-based~\cite{liu2019roberta}) fine-tuned using a 70\%-10\%-20\% train-validation-test split; and LLMs (GPT4.1-mini~\cite{gpt41} and Qwen3-30B~\cite{qwen3}) evaluated in a zero-shot setting, directly applying the prompt in \autoref{tab:promptAnnotation}. 
To address the imbalanced distribution of PTs, we apply techniques such as weighted loss functions and data resampling. 

\noindent\textbf{Classification Results.}
As shown in \autoref{tab:pt_classification},  RoBERTa-based model achieves the best overall performance, particularly excelling in recall ($89.87\%$) and F1-score ($86.69\%$). BERT-based model demonstrates comparable results, slightly behind RoBERTa but still robust. Traditional machine learning approaches, namely SVM and Random Forest, deliver surprisingly competitive performance. However, accuracy across all models remains relatively low, likely due to the inherent complexity and multi-label nature of the classification task. Notably, LLMs (GPT4.1-mini and Qwen3-30B) exhibit the worst performance, affirming the challenges of using prompt-based classification without human verification, as discussed in Section~\ref{subsec:dataannotation}.  


\subsection{Scam Completion}
\noindent\textbf{Evaluation Setting.} 
We evaluate this task using eight LLMs: six API-based models (GPT-4.1, GPT-4.1-mini, GPT-4o, Grok-3, Grok-3-mini, Gemini-2.0) and two open-source models (Qwen3-30B and Llama3-70B).
Given the beginning of the scam message, we first classify into topic, and select the most frequent PT combinations of the topic, feed them to LLM.
The prompt of this task is shown in \autoref{tab:promptCompletion} in \autoref{appendix:Prompt}.    
We experiment three input split settings ($20\%$, $40\%$, and $60\%$), which refer to how much of the original scam message is shown to the LLM. For example, a $20\%$ split gives an LLM only the first $20\%$ of the message and asks it to generate the rest. We then combine the input and the generated text to form the full generated scam message for evaluation.

\begin{figure}[t]
    \centering
    \includegraphics[width=1\columnwidth]{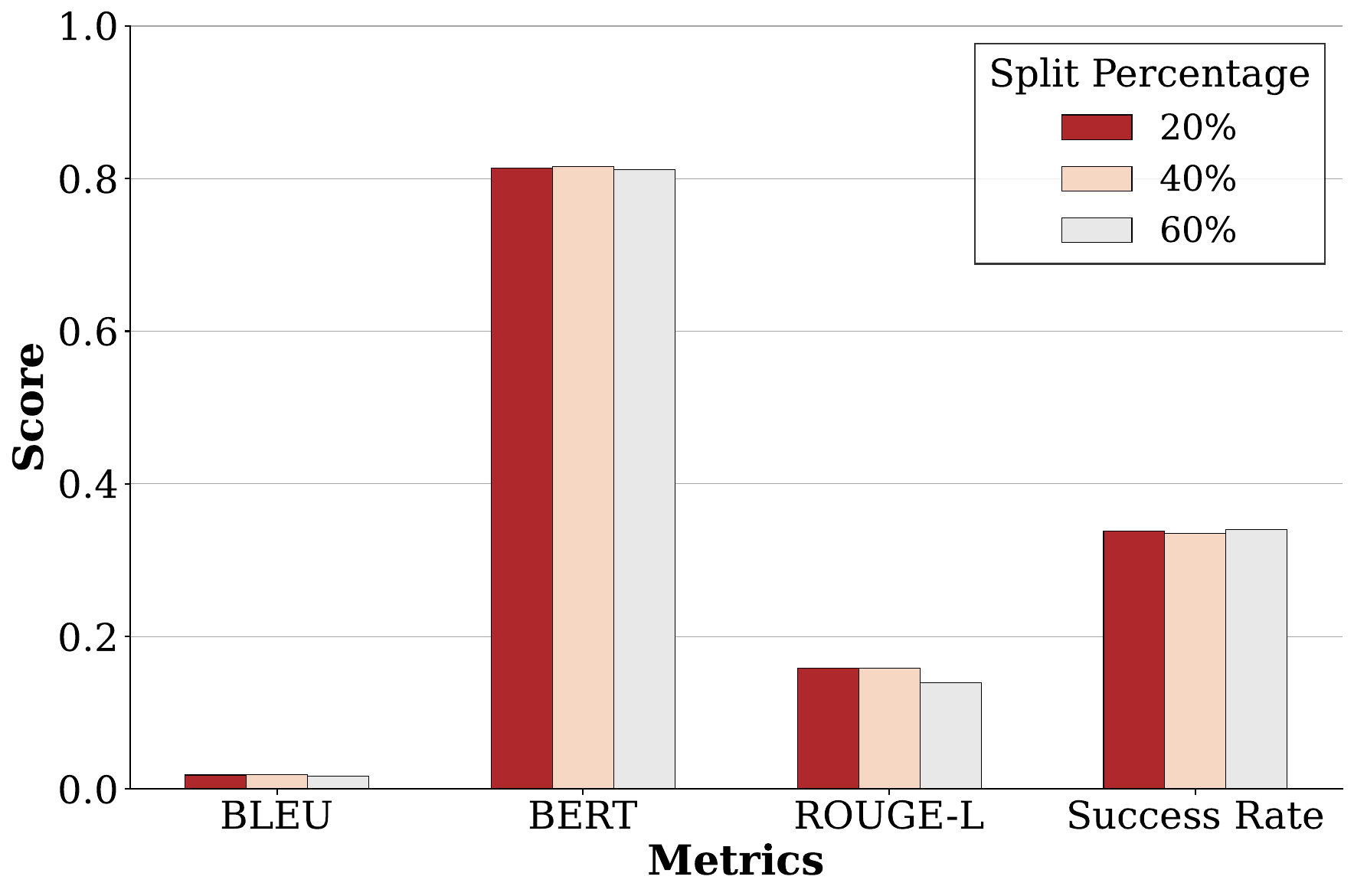}
    \caption{Impact of split percentages on scam completion performance.}
    \label{fig:splitpercentage}
\end{figure}

\begin{figure}[t]
    \centering
    \includegraphics[width=1\columnwidth]{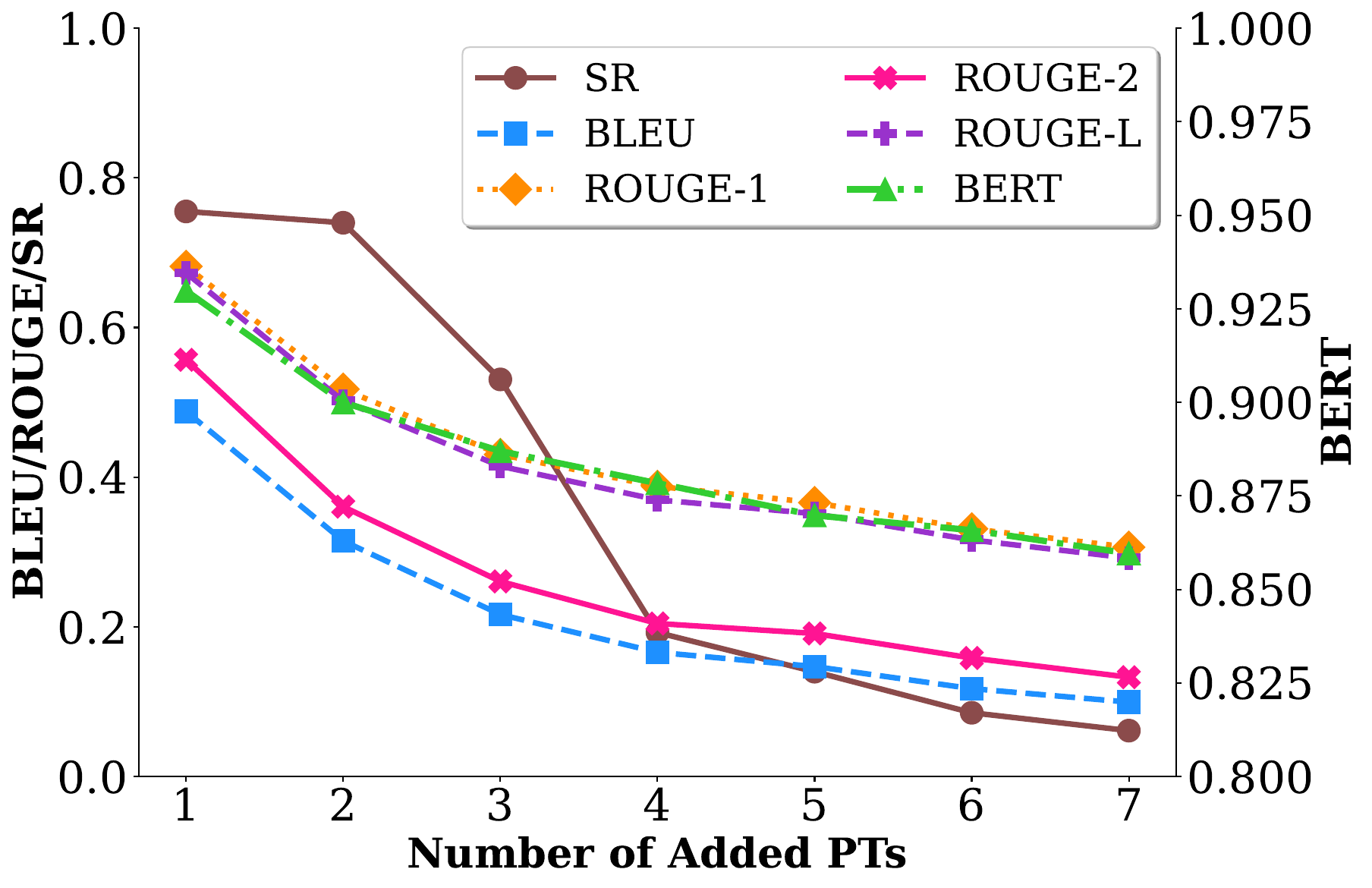}
    \caption{Impact of the number of generated PTs on scam augmentation performance.}
    \label{fig:numberPTs}
\end{figure}

\noindent\textbf{Results.}  
As illustrated in \autoref{tab:scam_generation}, all models demonstrate relatively low performance on ROUGE, BLEU, and BERTScore. Llama-3-70B achieves the highest ROUGE-1 ($0.0258$), Gemini-2.0 achieves the highest ROUGE-2 ($0.0269$) and GPT-4.1 leads in ROUGE-L ($0.1679$). GPT-4o obtains the best BLEU ($0.0219$) and BERTScore ($0.8181$).  
In contrast, the SR remains moderately better, approximately $0.3$ across all models (GPT-4.1-mini archives the highest $0.3376$).  
This suggests that while the generated text may differ in syntax and meaning, LLMs can still capture the conveyed PTs to a limited extent.

\noindent\textbf{Impact of Input Length.}
We further analyze how different input lengths (i.e., split settings) affects performance. 
We show the performance of GPT-4.1-mini, the model with the highest SR, in \autoref{fig:splitpercentage}.
It can be seen that all metrics show negligible variation across different split percentages. 
This indicates that simply increasing the input length does not improve the models' performance to incorporate the correct PTs.  One reason is that the generated text is often much longer than the original text, so metric scores are influenced more by the generated portion rather than the given input. A possible solution is to constrain the output length to match the original message.  


\subsection{Scam Augmentation}
\noindent\textbf{Evaluation Setting.} 
We evaluate this task using the same models employed in the Scam Completion task.

\noindent\textbf{Results.} 
As illustrated in \autoref{tab:scam_generation}, among the evaluated LLMs, Qwen3-30B notably achieves the highest performance for text generation metrics, including ROUGE-1 ($0.7867$), ROUGE-2 ($0.7253$), ROUGE-L ($0.7824$), BLEU ($0.6628$), and BERTScore ($0.9451$). Additionally, all models exhibit high SRs with GPT-4.1 achieving the highest at $0.8831$. 
Overall, all the models achieve better performance compared to the Scam Completion task. This is likely because, in this task, the full original scam message is provided, making it easier for LLMs to preserve tone, structure, and wording. 

\noindent\textbf{Impact of Added PTs.} 
We further investigate what causes the sharp performance gap between Scam Completion and Scam Augmentation.
We use GPT-4.1 (the model with the highest SR) and select messages in D2 that contain the very few PTs (2 PTs). We then modify the prompt (see \autoref{tab:promptAugmentation}) to ask the model to add 1 to 7 additional PTs. \autoref{fig:numberPTs} shows how metrics scores change as more PTs are added. 
We observe that as the number of added PTs increases, performance on all metrics declines, indicating the task becomes harder with each additional PT. This helps explain why Scam Completion results are generally worse than Scam Augmentation. On average, each message in D2 contains $3.53$ PTs, meaning Scam Completion implicitly requires the model to generate more PTs ($3.53$) from scratch, while Scam Augmentation typically requires adding only one. 

\begin{figure}[t]
    \centering
    \includegraphics[width=\columnwidth]{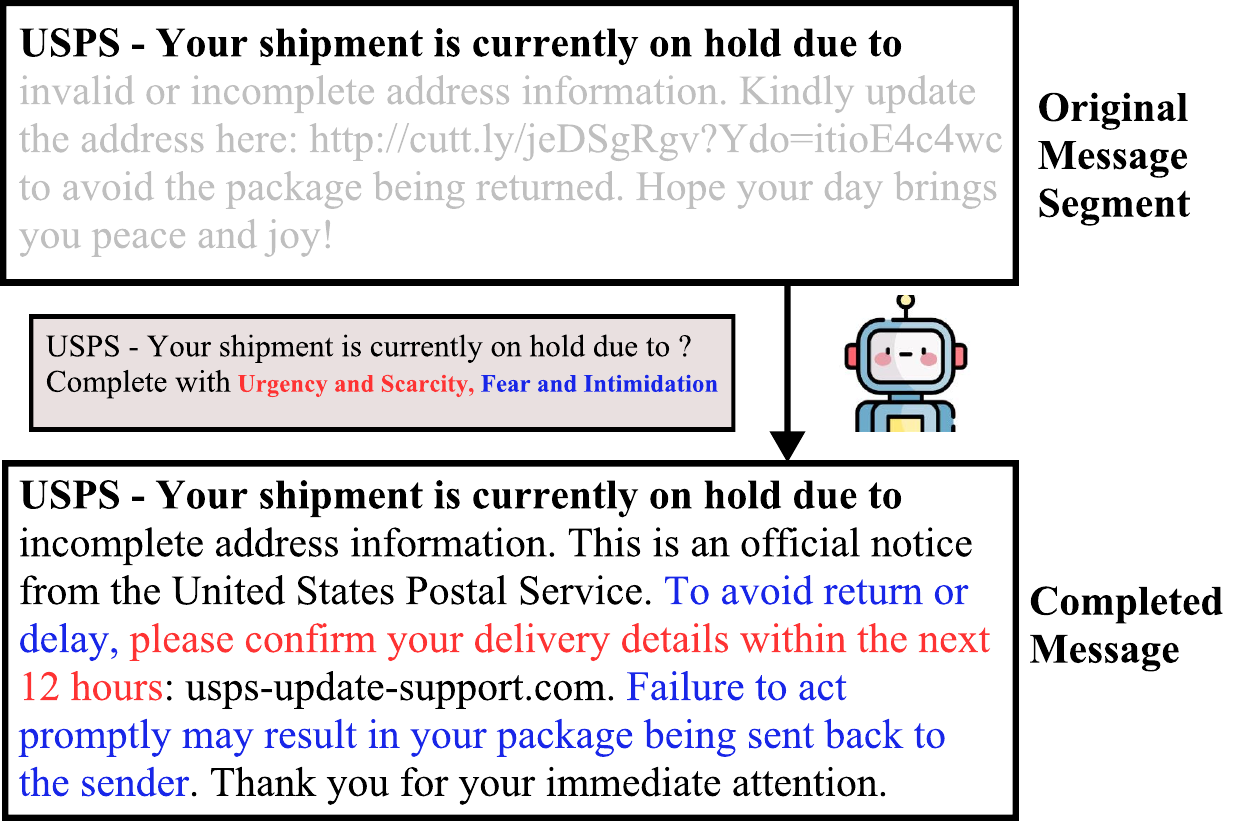}
    \caption{Case study: scam completion.}
    \label{fig:casestudyC}
\end{figure}
\begin{figure}[t]
    \centering
    \includegraphics[width=\columnwidth]{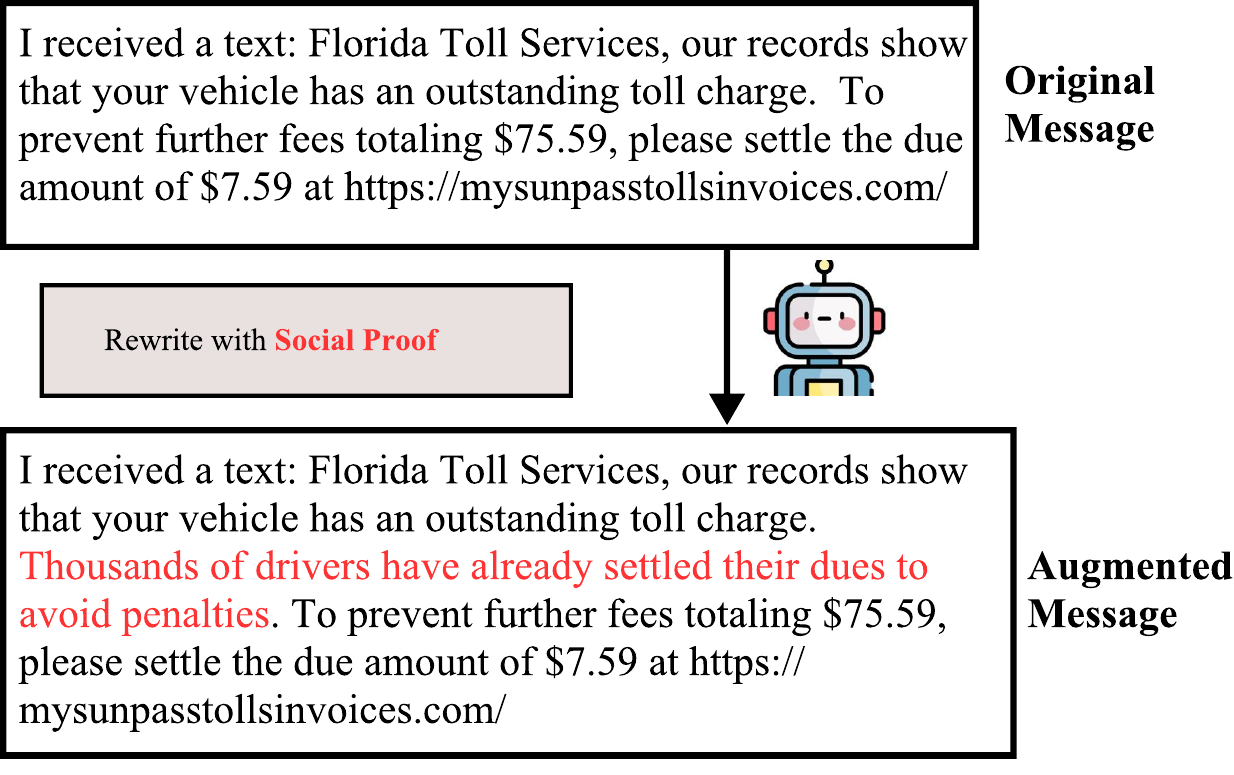}
    \caption{Case study: scam augmentation.}
    \label{fig:casestudyA}
\end{figure}

\subsection{Case Study}
To better understand the results of our evaluation, we conducted a case study analysis on the generated scam messages from both the Scam Completion and Augmentation tasks.  

\noindent\textbf{Limitations of N-gram Metrics.}
While n-gram metrics such as ROUGE, BLEU provide convenient approximations of text similarity, they often fail to capture semantic similarity. Through manual inspection, we observe that many generated scam messages successfully convey the intended PT using alternative phrasing and varied sentence structures. For example, in the case study shown in \autoref{fig:casestudyC}, the generated message effectively conveys the original PTs by exploiting  ``Urgency and Scarcity'' and ``Fear and Intimidation'' without directly reusing phrases from the original message. This demonstrates the capability of LLMs to generate creatively diverse yet semantically aligned scam completions.

\noindent\textbf{LLM Hallucination.} 
Despite strong quantitative results in the Scam Augmentation task, we also observe instances where LLM introduces unnatural expressions within the scam context when integrating PTs. For example, in the unpaid toll-fee scam illustrated in \autoref{fig:casestudyA}, the augmented version of an unpaid toll-fee scam incorporates the ``Social Proof'' technique using the sentence: \textit{Thousands of drivers have already settled their dues to avoid penalties''}. While the PT is present, the phrasing feels unnaturally desperate and may raise user suspicion. A more contextually appropriate revision, such as \textit{``Many drivers incur additional charges when payment is delayed''} would better preserve the tone and subtlety typical of real scam messages. This suggests that while LLMs effectively generate the specified PT, their outputs may require contextual adjustment to ensure realism and credibility of the scam message.

\section{Discussion and Future Work}


\noindent\textbf{Imbalanced Distribution of PTs.}  
\blue{
\autoref{fig:ptdistribution} illustrates the imbalanced distribution of PTs in our dataset.  
Although this distribution reflects real-world scam patterns, it negatively impacts model performance.    
To address this issue, we adopt two strategies. First, we propose the Scam Augmentation task that can generate additional examples of underrepresented PTs. Second, we apply techniques such as weighted loss functions and data resampling during PT classification.  
For future work, enlarging the dataset (e.g., by labeling the full D0) may provide more examples of scarce PTs and further improve model's robustness. }


\noindent\textbf{Realism of LLM-Generated Scams.}  
\blue{ While automatic metrics such as BERTScore and BLEU provide useful signals, they cannot fully capture the realism and persuasiveness of LLM-generated scams.  In future work, we plan to conduct a comprehensive, large-scale human study under an IRB-approved protocol. Human annotators will systematically evaluate generated scams for coherence, adherence to specified PTs, and overall realism, and their feedback can be leveraged to further refine LLM-based generation.
}

\noindent\textbf{Scam Incident Association via Psychological Patterns.}
Scam incidents often occur in spikes, with multiple cases emerging over a short period of time that share similar tactics but differ in surface-level content. For example, toll road scams reported across various U.S. states, such as Arizona, Florida, California, Washington, and Ohio~\cite{azdot,sunpass,fastrak,ezpass,wsdot,ohdot,quickpass}, exhibit different text formats and sender names but consistently exploit the same psychological techniques, such as \textit{Authority and Impersonation} and \textit{Urgency and Scarcity}, as illustrated in \autoref{fig:goldenexample}. These recurring behavioral patterns parallel cyber attacks that reuse the same technical tactics and techniques~\cite{attck}, suggesting that PTs can serve as behavioral signatures to associate and cluster scam incidents. Future work could develop PT-based clustering or temporal analysis methods to automatically link related scams, enabling earlier detection, trend analysis, and coordinated response to emerging scam campaigns.


\noindent\textbf{Leveraging Victim Narratives for Scam Explanation.}
Beyond scammer messages, many scam reports include rich victim narratives that describe how individuals recognized the scam or were manipulated step by step (e.g., \autoref{fig:scam_report_narrative}). These narratives contain cognitive and emotional processes of victims, which resemble the reasoning process of LLMs. Future research could leverage this dimension to develop scam detection systems that model both the attacker’s persuasive tactics and the victim’s reaction. Such models may enable more interpretable scam alerts or personalized warnings based on user susceptibility.

\noindent\textbf{Toward Real-Time Detection and Prevention.}   
Our findings show that LLMs can detect and predict scam content even from partial inputs. This opens avenues for real-time scam detection systems that operate on incomplete or unfolding messages. However, ensuring robustness, reducing hallucinations, and maintaining natural tone in generation remain open challenges. Future work could explore fine-tuning techniques to better align LLM outputs with real-world scam characteristics.

\section{Conclusion}
In this work, we introduce \bench, the first benchmark designed to systematically capture and evaluate PTs  in real-world scam incidents. By collecting diverse scam reports from six public reporting platforms and grounding our annotations in established cognitive and persuasion theories, \bench~bridges the gap between  psychology and practical cyber security analysis. Our human–LLM collaborative annotation framework enables scalable, high-quality PT labeling and our evaluation on three downstream tasks shows that \bench~poses challenges to existing models. We believe \bench~lays the foundation for future research on scam detection and generation, persuasive language understanding, and the development of trustworthy AI systems for combating online scams and fraud.

\section*{Limitations}


This work has several limitations.

First, the taxonomy of PTs presented in \autoref{tab:PTs} is manually constructed based on established psychological theories and a preliminary study of real-world scam reports. While it captures a wide range of manipulation strategies commonly observed in scams, it may not fully encompass the entire psychological landscape in scams. For instance, techniques such as enforced isolation~\cite{lea2009psychology,scamIsolation} where victims are instructed not to disclose the situation to others (e.g., ``They also asked us not to tell anyone about this'') but are not explicitly included in our current taxonomy. Future work could consider expanding the PT framework to account for more PTs.

Second, our prompting strategy, while effective, offers ample room for refinement. The current few-shot prompts include limited examples for each PT, which may restrict the LLM’s ability to generalize to ambiguous or borderline cases. Future improvements could include richer in-context demonstrations, dynamically selected examples, or fine-tuning to enhance LLM reasoning and reduce false positives.

Finally, while \bench~includes diverse scams across multiple platforms and regions, all reports are currently in English. Given that scams are a global issue, extending \bench~to include reports written in other major languages, such as Chinese, Spanish, or Arabic, would be critical for broader applicability and cross-cultural analysis. This would also enable benchmarking multilingual scam detection systems and studying language-specific variations in scammer persuasion strategies.

\section*{Ethics Statement}
All datasets used in our study are publicly accessible through their respective websites. We did not violate any licenses in the process of downloading data. Our use of these datasets is strictly for academic purposes.

This study explores the capabilities of LLMs in generating scam content for research purposes. While our experiments involve generating scam-like messages, all experiments are conducted in a controlled setting strictly for defensive research and evaluation. We emphasize that our methodology is intended to support the development of scam detection systems and raise awareness of potential misuse.

Notably, prior work~\cite{roy2024chatbots} has demonstrated that commercial LLMs can be prompted to generate scam websites and emails. In our study, we evaluate both open-source and commercial models, including OpenAI and Grok. We observe that only OpenAI’s most recent reasoning models (GPT-o3 and GPT-o4) consistently refuse to generate scam content, while others do not implement similar safeguards. This underscores the importance of integrating robust content filtering mechanisms into generative models. We strongly advocate for responsible AI development and stress that all findings in this paper are presented solely to enhance understanding and strengthen fraud prevention efforts.

\section*{Acknowledgments}

The work of S. Ma, T. Ma, and Y. Ye was partially supported by the NSF under grants IIS-2533550, IIS-2321504, IIS-2340346, IIS-2217239, CNS-2426514, and CMMI-2146076, ND-IBM Tech Ethics Lab Program, Notre Dame Strategic Framework Research Grant (2025), and Notre Dame Poverty Research Package (2025). 
Xusheng Xiao's work is partially supported by the National Science Foundation under the grants CCF-2318483 and CNS-2438197. The work of Jiahao Liu and Zhenkai Liang is partially supported by the National Research Foundation, Singapore under its Industry Alignment Fund – Pre-positioning (IAF-PP) Funding Initiative. 
Any expressed opinions, findings, and conclusions or recommendations are those of the authors and do not necessarily reflect the views of National Research Foundation, Singapore and other sponsors.


\bibliography{emnlp2023}
\bibliographystyle{acl_natbib}

\newpage
\appendix
\label{sec:appendix}

\section{Additional Figures and Tables}
\label{appendix:Prompt}

\begin{figure}[h]
    \centering
    \includegraphics[width=\columnwidth]{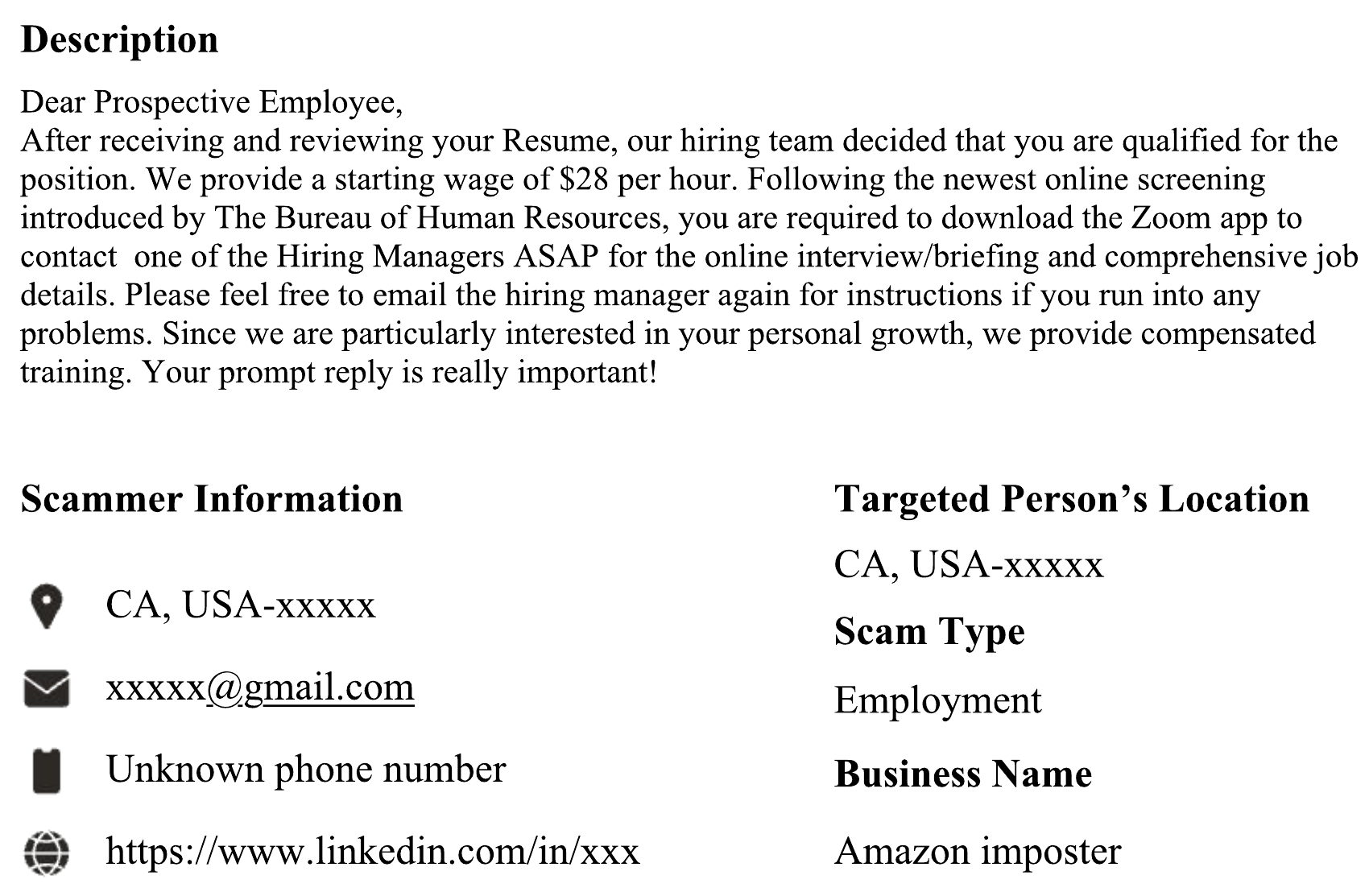}
    \caption{A typical scam report.}
    \label{fig:scam_report}
\end{figure}

\begin{figure}[h]
    \centering
    \includegraphics[width=\columnwidth]{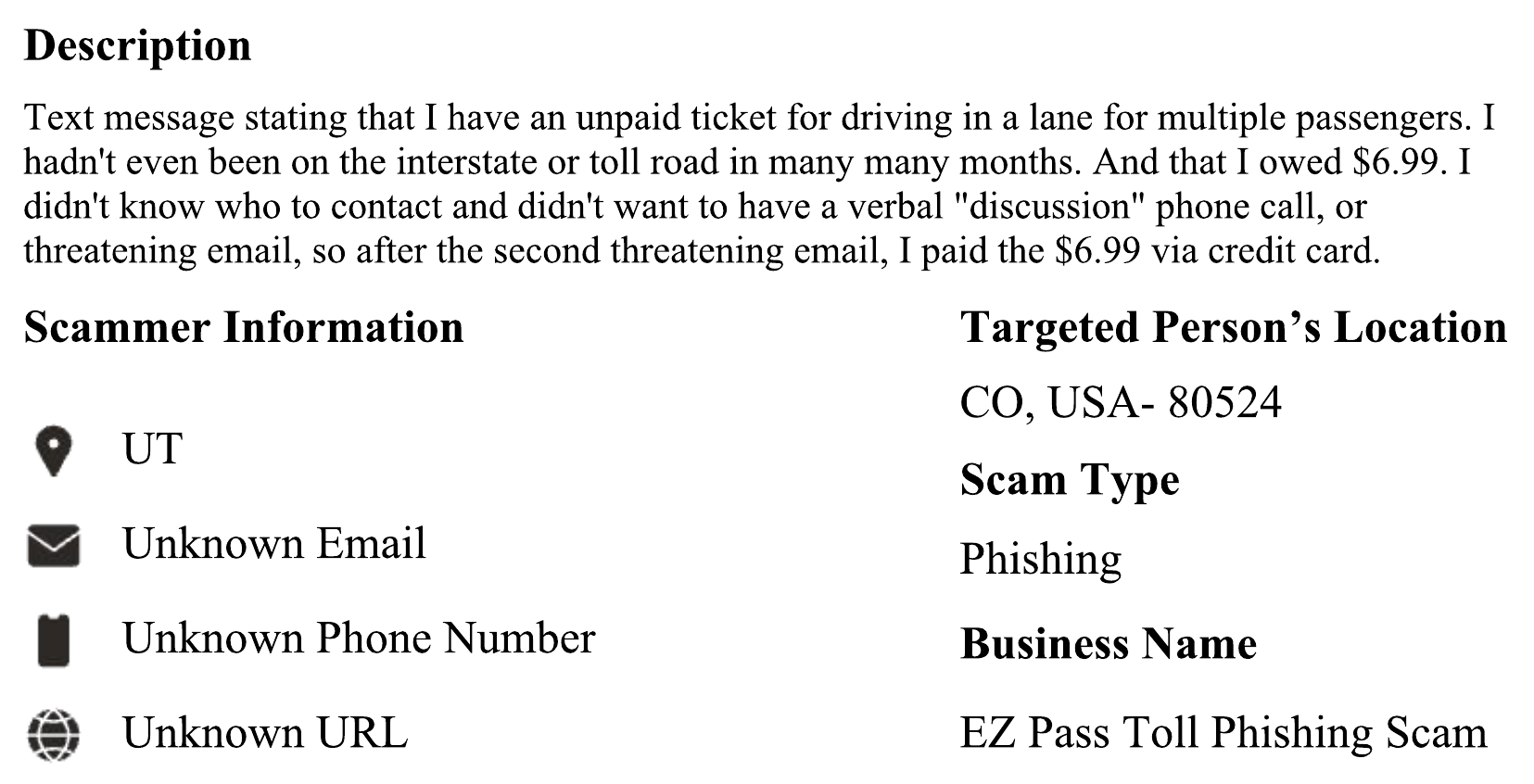}
    \caption{Scam report example: victim narrative~\cite{scamReportVictimNarrative}.}
    \label{fig:scam_report_narrative}
\end{figure}

\begin{figure}[h]
    \centering
    \includegraphics[width=\columnwidth]{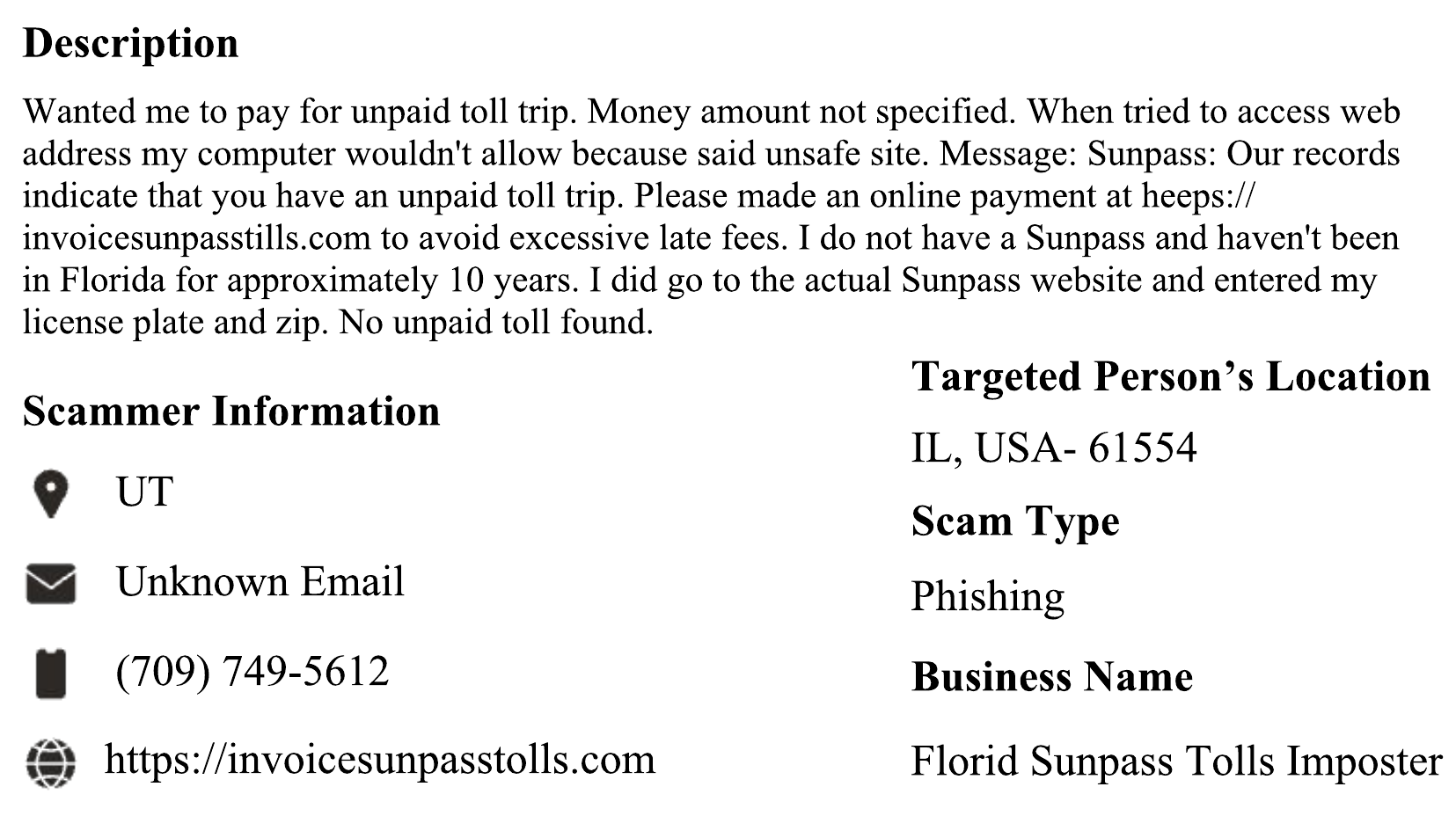}
    \caption{Scam report example: victim narrative and scam message~\cite{scamReportMix}.}
    \label{fig:scam_report_mix}
\end{figure}

\begin{table*}[h]
\resizebox{\linewidth}{!}{%
\begin{tabular}{|c|c|l|}
\hline
\textbf{Role}         & \textbf{Category} & \multicolumn{1}{c|}{\textbf{Content}}                                                                                                                                                                                                                                                                                                                                                                                                                                                                                                                                                                                                                                                                                                                                                                                                                                                                                                                                                                                                                                                                                                                                                                                                                                                                                                                                                 \\ \hline
System                & -                 & \begin{tabular}[c]{@{}l@{}}Scammers use social engineering attacks that exploit psychological techniques \\ to manipulate victims. Our goal: We collect a number of scam reports and \\ aim to extract the psychological techniques used in these scam reports.\end{tabular}                                                                                                                                                                                                                                                                                                                                                                                                                                                                                                                                                                                                                                                                                                                                                                                                                                                                                                                                                                                                                                                                                                          \\ \hline
\multirow{3}{*}{User} & Task description  & \begin{tabular}[c]{@{}l@{}}Now I give you the victim report. \\ Please extract any psychological techniques exploited by the scammer.\\ Requirement 1: if no psychological techniques is identified, return an empty dictionary:\\ \{\} \\ Avoid guess, you must return the psychological techniques when you are prettry sure it exists \\ Requirement 2: Return the output strictly as a JSON dictionary:\\ \{"psychological technique A": Corresponding texts in the victim report,\\ "psychological technique B": Corresponding texts in the victim report,...\} \\ For example: \\ \{'Reciprocity': "This is a work fromlly plan your own day schedule ...", \\  'Consistency': "We received your application for our Remote Customer Enrollment Position ..."\}\end{tabular}                                                                                                                                                                                                                                                                                                                                                                                                                                                                                                                                                                                                    \\ \cline{2-3} 
                      & Scam message      & “USPS - Your shipment is currently..."                                                                                                                                                                                                                                                                                                                                                                                                                                                                                                                                                                                                                                                                                                                                                                                                                                                                                                                                                                                                                                                                                                                                                                                                                                                                                                                                                \\ \cline{2-3} 
                      & PT definition     & \begin{tabular}[c]{@{}l@{}}We consider following psychological techniques:\\ -Authority and Impersonation: From Cialdini’s 6 principles of persuasion:\\ people tend to obey authorities. People trust credible individuals. \\ -Phantom Riches: Visceral triggers of desire that override rationality \\ -Fear and Intimidation: Leverages the fear response which \\ overrides rational thought \\ -Liking: From Cialdini’s 6 principles of persuasion, \\ Preference for saying “yes” to the requests of people they know and like. \\ People are programmed to like others who like them back and \\ who are similar to them. \\ -Urgency and Scarcity: From Cialdini’s 6 principles of persuasion, \\ sense of urgency and scarcity assign more value to items \\ -Pretext and Trust: Scammer make up story to add source credibility \\ and gain victim's trust. \\ -Reciprocity: From Cialdini’s 6 principles of persuasion. \\ Tendency to feel obliged to repay favors from others. \\ “I do something for you, you do something for me.” \\ -Consistency: From Cialdini’s 6 principles of persuasion. \\ Tendency to behave in a way consistent with past decisions and behaviors \\ -Social Proof: From Cialdini’s 6 principles of persuasion. \\ Tendency to reference the behavior of others, \\ by using the majority behavior to guide their own actions.\end{tabular} \\ \hline
\end{tabular}
}
\caption{The prompt for LLM annotation.}
\label{tab:promptAnnotation}
\end{table*}

\begin{table*}[h]
\resizebox{\linewidth}{!}{%
\begin{tabular}{|c|c|l|}
\hline
\textbf{Role}         & \textbf{Category} & \multicolumn{1}{c|}{\textbf{Content}}                                                                                                                                                                                                                                                                                                                                                                                                                                                                                                                                                                                                                                                                                                                                                                                                                                                                                                                                                                                                                                                                                                                                                                                                                                                                                                                                                    \\ \hline
System                & -                 & \begin{tabular}[c]{@{}l@{}}You are an assistant that completes scam messages incorporating \\ specific psychological techniques.\end{tabular}                                                                                                                                                                                                                                                                                                                                                                                                                                                                                                                                                                                                                                                                                                                                                                                                                                                                                                                                                                                                                                                                                                                                                                                                                                            \\ \hline
\multirow{3}{*}{User} & Task description  & \begin{tabular}[c]{@{}l@{}}You are tasked with completing a scam message based on its beginning. \\ The message should incorporate specific psychological techniques. \\ Ensure your completion continues directly from the last word of the\\  provided beginning, maintaining the same style and tone. Only return \\ the completion without any additional text or explanation.\end{tabular}                                                                                                                                                                                                                                                                                                                                                                                                                                                                                                                                                                                                                                                                                                                                                                                                                                                                                                                                                                                          \\ \cline{2-3} 
                      & Scam message      & \begin{tabular}[c]{@{}l@{}}Beginning of the message:\\ “Text message : USPS - Your shipment is currently\end{tabular}                                                                                                                                                                                                                                                                                                                                                                                                                                                                                                                                                                                                                                                                                                                                                                                                                                                                                                                                                                                                                                                                                                                                                                                                                                                                    \\ \cline{2-3} 
                      & PT definition     & \begin{tabular}[c]{@{}l@{}}We consider following psychological techniques:\\ -Authority and Impersonation: From Cialdini’s 6 principles of persuasion:\\ people tend to obey authorities. People trust credible individuals. \\ -Phantom Riches: Visceral triggers of desire that override rationality \\ -Fear and Intimidation: Leverages the fear response which \\ overrides rational thought \\ -Liking: From Cialdini’s 6 principles of persuasion, \\ Preference for saying “yes” to the requests of people they know and like. \\ People are programmed to like others who like them back and \\ who are similar to them. \\ -Urgency and Scarcity: From Cialdini’s 6 principles of persuasion, \\ sense of urgency and scarcity assign more value to items \\ -Pretext and trust:  Scammer make up story to add source credibility \\ and gain victim's trust.  \\ -Reciprocity: From Cialdini’s 6 principles of persuasion. \\ Tendency to feel obliged to repay favors from others. \\ “I do something for you, you do something for me.” \\ -Consistency: From Cialdini’s 6 principles of persuasion. \\ Tendency to behave in a way consistent  with past decisions and behaviors \\ -Social Proof: From Cialdini’s 6 principles of persuasion. \\ Tendency to reference the behavior of others, \\ by using the majority behavior to guide their own actions.\end{tabular} \\ \hline
\end{tabular}
}
\caption{The prompt for the scam completion task.}
\label{tab:promptCompletion}
\end{table*}

\begin{table*}[h]
\resizebox{\linewidth}{!}{%
\begin{tabular}{|c|c|l|}
\hline
\textbf{Role}         & \textbf{Category} & \multicolumn{1}{c|}{\textbf{Content}}                                                                                                                                                                                                                                                                                                                                                                                                                                                                                                                                                                                                                                                                                                                                                                                                                                                                                                                                                                                                                                                                                                                                                                                                                                                                                                                                                 \\ \hline
System                & -                 & \begin{tabular}[c]{@{}l@{}}You are an assistant that rewrites scam messages incorporating\\ specific psychological techniques\end{tabular}                                                                                                                                                                                                                                                                                                                                                                                                                                                                                                                                                                                                                                                                                                                                                                                                                                                                                                                                                                                                                                                                                                                                                                                                                                            \\ \hline
\multirow{3}{*}{User} & Task description  & \begin{tabular}[c]{@{}l@{}}Please rewrite the scam message to also include the following psychological technique:\\ {PT Name}: {PT Definition}\\ Make sure to keep all the original facts intact while incorporating this new PT.\\ Only return the rewritten message without any additional text or explanation.\end{tabular}                                                                                                                                                                                                                                                                                                                                                                                                                                                                                                                                                                                                                                                                                                                                                                                                                                                                                                                                                                                                                                                        \\ \cline{2-3} 
                      & Scam message      & \begin{tabular}[c]{@{}l@{}}The scam message:\\ “USPS - Your shipment is currently...\end{tabular}                                                                                                                                                                                                                                                                                                                                                                                                                                                                                                                                                                                                                                                                                                                                                                                                                                                                                                                                                                                                                                                                                                                                                                                                                                                                                     \\ \cline{2-3} 
                      & PT definition     & \begin{tabular}[c]{@{}l@{}}We consider following psychological techniques:\\ -Authority and Impersonation: From Cialdini’s 6 principles of persuasion:\\ people tend to obey authorities. People trust credible individuals. \\ -Phantom Riches: Visceral triggers of desire that override rationality \\ -Fear and Intimidation: Leverages the fear response which \\ overrides rational thought \\ -Liking: From Cialdini’s 6 principles of persuasion, \\ Preference for saying “yes” to the requests of people they know and like. \\ People are programmed to like others who like them back and \\ who are similar to them. \\ -Urgency and Scarcity: From Cialdini’s 6 principles of persuasion, \\ sense of urgency and scarcity assign more value to items \\ -Pretext and trust: Scammer make up story to add source credibility \\ and gain victim's trust. \\ -Reciprocity: From Cialdini’s 6 principles of persuasion. \\ Tendency to feel obliged to repay favors from others. \\ “I do something for you, you do something for me.” \\ -Consistency: From Cialdini’s 6 principles of persuasion. \\ Tendency to behave in a way consistent with past decisions and behaviors \\ -Social Proof: From Cialdini’s 6 principles of persuasion. \\ Tendency to reference the behavior of others, \\ by using the majority behavior to guide their own actions.\end{tabular} \\ \hline
\end{tabular}
}
\caption{The prompt for the scam augmentation task.}
\label{tab:promptAugmentation}
\end{table*}

\end{document}